\documentclass[manuscript]{raa_rb}
\usepackage{graphicx,times}
\usepackage{natbib}
\usepackage{amssymb,amsmath}
\usepackage{enumerate}
\bibpunct{(}{)}{;}{a}{}{,}

\usepackage[a4paper=true,pagebackref=true]{hyperref}
\hypersetup{pdftitle = The title of my PDF, pdfauthor = My name, pdfsubject= The subject, pdfkeywords = keyword1 keyword2 keyword3} 

\begin{document}

   \title{Ionospheric inversion of the Venus Express radio occultation data observed by Shanghai 25 m and New Norcia 35 m antennas 
$^*$
\footnotetext{\small $*$ Supported by the National Natural Science Foundation of China (Grant No. 11103063, 11178008), the national key basic research and development plan (Grant No. 2015CB857101), and partly supported by the Key Laboratory of Planetary Sciences (Grant No. PSL15\_04).}
}

 \volnopage{ {\bf 2014} Vol.\ {\bf X} No. {\bf XX}, 000--000}
   \setcounter{page}{1}

   \author{Su-jun Zhang\inst{1,2,3}, Nian-chuan Jian\inst{1},  Jin-ling Li\inst{1}, Jin-song Ping\inst{4}, Cong-yan Chen\inst{5}, Ke-fei Zhang
      \inst{6}
   }

   \institute{ Shanghai Astronomical Observatories, Chinese Academy of Sciences,
             Shanghai 200030, China; {\it sjzhang@shao.ac.cn}\\
         \and 
            Key Laboratory of Planetary Sciences, Chinese Academy of Sciences, Shanghai, 200030, China\\       
        \and 
            University of Chinese Academy of Sciences, Beijing 100049, China\\
        \and
             Key Laboratory of Lunar and Deep Space Exploration, National Astronomical Observatories, Chinese Academy of Sciences,
             Beijing 100012, China \\
        \and
            School of Information Science and Engineering, Southeast University, Nanjing 210096, China\\
        \and
             RMIT University, Melbourne Victoria 3001, Australia\\
\vs \no
   {\small Received 2014 ; accepted 2015}
}

\abstract{Electron density profiles of the Venus' ionosphere are inverted from the Venus Express (VEX) one-way open-loop radio occultation experiments carried out by Shanghai 25 m antenna from November 2011 to January 2012 at solar maximum conditions and by New Norcia 35 m antenna from August 2006 to June 2008 at  solar intermediate conditions. The electron density profile (from 110 km to 400 km) retrieved from the X-band egress observation at Shanghai station, shows a single peak near 147 km with a peak density of about $2 \times 10^4 \rm{cm}^{-3}$ at a solar zenith angle of 94$^{\circ}$.  As a comparison, the VEX radio science (VeRa) observations at New Norcia station were also examined, including S-, X-band and dual-frequency data in the ingress mode. The results show that the electron density profiles retrieved from the S-band data are more analogous to the dual-frequency data in the profile shape, compared with the X-band data. Generally, the S-band results slightly underestimate the magnitude of the peak density, while the X-band results overestimate that. The discrepancy in the X-band profile is probably due to the relatively larger unmodeled orbital errors. It is also expected that the ionopause height is sensitive to the solar wind dynamical pressure in high and intermediate solar activities, usually in the range of 200 km - 1000 km on the dayside and much higher on the nightside. Structural variations (``bulges'' and fluctuations) can be found in the electron density profiles in intermediate solar activity, which may be caused by the interaction of the solar wind with the ionosphere. Considerable ionizations can be observed in the Venus' nightside ionosphere, which are unexpected for the Martian nightside ionosphere in most cases. 
}

   \authorrunning{S.-J. Zhang et al. }            
   \titlerunning{ Ionospheric inversion of the Venus Express radio occultation data}  
   \maketitle

%
\section{Introduction}           

The dayside ionosphere of Venus is produced locally by the photoionizaiton of the solar EUV and soft X-ray radiations along with impact ionization by photoelectrons and secondary electrons, while the nightside ionosphere is produced by a combination of ion flow from the dayside and local ion production by the suprathermal electron impact ionization \citep[e.g.,][]{Zhang1990,Fox2011}. The super rotation of the Venus' atmosphere and the lack of an intrinsic magnetic field make the nightward ion flow play an important role in the formation of the nightside ionosphere \citep{Russell1980}. The magnetic field of the Venus' ionosphere is induced by the interaction between its ionosphere and the solar wind, and can also be viewed as a compression of the interplanetary magnetic field (IMF) as it drapes around the ionosphere. The weak magnetic field of Venus can provide negligible protection to the atmosphere against the solar radiation \citep{Brace1991}.

The electron density profiles returned from the Pioneer Viking Orbiter (PVO) radio occultation (RO) observations show that the nightside ionosphere of Venus exists regardless of the solar activity, most of time with a robust density peak \citep{Kliore1992}. In contrast, the Martian nightside ionosphere is a sporadic phenomenon, the peak densities are weak in most conditions or even don't exist at all \citep{Zhang1990,Kliore1992}. The nightside ionosphere of Venus is also highly variable, especially in the regions near and above the peak altitude, mainly due to the variations of the number of ions transported from the dayside which is correlated with the solar fluxes, and the altitude of the ionopause which is anticorrelated with the solar wind dynamic pressure \citep{Cravens1981a,Fox2011}.

An ionopause can be formed between the solar wind plasma and the ionospheric plasma, where the external pressures ( the solar wind dynamic pressure, its thermal pressure and the magnetic pressure) are balanced with the internal pressures ( the ionospheric thermal pressure and the field pressure) \citep{Brace1991,Luhmann1991}. The ionopause height can be defined as the altitude where the electron or ion density passes through a value of $10^{2}$ cm$^{-3}$ in a steep gradient as for the Langmuir probe and the retarding potential analyzer experiments \citep{Brace1983,Knudsen1979}, or the altitude where the electron density first falls below $5 \times 10^{2}$ cm$^{-3}$ for the RO observations \citep{Kliore1991,Kliore1992}, or the boundary where the magnetic pressure transforms to thermal pressure as for the magnetometer experiments \citep{Phillips1984}, or the boundary between the thermal and suprathermal ion components as for the ion mass spectrometer experiments \citep{Taylor1980}. During the solar maximum, the plasma pressure exceeds the solar wind pressure to form a high ionopause, the ionization transport from the dayside dominates in this condition. During the solar minimum, the ionopause is much lower, which may prevent the dayside transport of ions leaving only the contribution of impact ionization by energetic electrons \citep{Kliore1992}.  

Radio occultation is one of the most important techniques to explore the planetary atmosphere and ionosphere, which utilizes the radio links between the spacecraft around the target planet and the antenna on Earth. The ionosphere of Venus was firstly detected by the radio occultation experiment of Mariner 5 in 1967 \citep{Mariner1967,Kliore1967}, and the subsequent Venera, Mariner, PVO and Magellan programs \citep{Fjeldbo1975,Kholodny1979,Knudsen1979,Jenkins1994}. In addition to the remote sensing experiments, the in-situ instruments mounted on Venera and  PVO (from 1978 -1992) spacecrafts had also measured the Venus' ionosphere for 6 and over 12 years, respectively. The large body of data from both in-situ and RO measurements allow extensive studies on the structure and temporal behavior of the Venus' ionosphere, as well as the comparisons with the Martian ionosphere \citep{Brace1991,Kliore1990,Kliore1991,Kliore1992,Luhmann1992,Knudsen1992}. 

There were many efforts on the theoretical modeling of the Venus' dayside ionosphere \citep{Nagy1980,Cravens1980,Kim1989,Shinagawa1988,Shinagawa1993,Shinagawa1996a,Shinagawa1996b}, the studies on the near-terminator and nightside ionosphere of Venus \citep{Fox1992,Fox2011,Fox2007,Luhmann1982,Mahajan2001}, and studies on the solar wind interaction with the Venus' ionosphere \citep{Taylor1980,Russell2006,Terada2004,Terada2009}. \citet{Cravens1981a,Cravens1981b} interpreted the behavior of the ionospheric peak on the dayside from electron density profiles from the PVO RO observations by comparison with model results.  \citet{Brace1983} found the exist of large amplitude post-terminator waves structures in the electron density and electron temperature profiles below 175 km at SZAs between $90^{\circ}$ and $120^{\circ}$. Ionospheric holes or plasma depletions in electron and ions were found in the nightside ionosphere \citep{Brace1980,Brace1982,Taylor1980}, and the production mechanism was also studied by \citet{Grebowsky1981} and \citet{Grebowsky1983}. 

The Venus Express (VEX) spacecraft is the latest Venus exploration mission after the PVO and Magellan program, and is the first European mission to planet Venus. The main objective of the VEX mission is to investigate the atmosphere and plasma environment of Venus from a polar orbit when the spacecraft is occultated observing from the ground station, and also aspects of the geology and surface physics in a comprehensive way \citep{Svedhem2007}. The complete descriptions of the VEX radio science (VeRa)  experiments can be found in \citet{Hausler2006}.

From the VEX RO observations, \citet{Patzold2007} discussed the day-to-day changes in the Venus' ionosphere from the radio sounding data of the first VeRa occultation season; \citet{Patzold2009} identified a sporadic layer of meteoric origin in the Venus lower ionosphere; \citet{Peter2014} compared the electron density profiles retrieved from the VeRa observations with those simulated from a one-dimensional photochemical model. Independent VEX RO experiments in China were also carried out with the Shanghai 25 m antenna. This paper mainly deals with the ionosphere inversion from different occultation modes and the structural variations revealed from the VEX RO data collected at Shanghai and New Norcia antennas.

This paper is structured as follows: Section 1 introduces the backgrounds of the Venus' ionosphere. Section 2 deals with the VEX radio sounding experiments of the Venus' ionosphere. Section 3 discusses the retrieval of the ionospheric parameters. Sections 4 gives the results retrieved from the RO observations at Shanghai and New Norcia stations. Section 5 discusses the variation of the ionopause altitude revealed from the electron density profiles. Section 6 concludes this paper.

\section{Radio sounding of the Venus' ionosphere}
The VEX spacecraft was launched on 9 November, 2005 and arrived Venus on 11 April, 2006. The scientific payloads on VEX are inherited from the Mars Express (MEX) and Rosetta spacecrafts, which permit direct comparisons of different planets due to the same instrument errors \citep{Hausler2006}. Two coherent one-way radio signals (S-band at 2.3 GHz and X-band at 8.4 GHz) are used to investigate the Venus' surface, neutral atmosphere, ionosphere and gravity field. The Ultrastable Oscillator (USO) installed on VEX is a direct derivative of the Rosetta's USO, with an Allan deviation of $\sim 3 \times 10^{-13}$ at 1-100 s. The high stability of the onboard USO guarantees the egress occultation can be conducted successfully, as the downlink signal is controlled by the reference signal on the spacecraft in this mode \citep{Hausler2006}. Meanwhile, the coherent downlink signals allow the separation of dispersive media effects from the classical Doppler shifts. 

During the 12th occultation season of the VEX spacecraft, several radio occultation experiments were conducted by the Shanghai 25 m antenna. The 600 MHz intermediate frequency (IF) radio signal is recorded digitally by a Radio Science Receiver (RSR), which is developed jointly by Southeast University and Shanghai Astronomical Observatory. The IF signal is down converted and desampled to a baseband signal ($\sim$~200 KHz), where the Doppler shift is computed via the self-developed frequency estimation scheme.  After subtracting the classical geometrical Doppler shift caused by the relative movement between the spacecraft and the ground station, the atmosphere and ionosphere media effect of Earth, signal variations caused by the Venus' atmosphere and ionosphere are reserved in the Doppler residuals. Then the Doppler residuals can be used to retrieve the molecular number density, pressure and temperature profiles of the atmosphere and electron density profiles of the ionosphere by the planetary occultation observation processing software, which is described in \citet{Zhang2011}. 

The vertical resolution of the RO experiment is determined by the radius of the first Fresnel zone, in the form of $(\lambda D)^{1/2}$, where $\lambda$ is the wavelength of transmitting signal, $D$ is the distance from the transmitter of the spacecraft to the closest approach of the ray path to the limb of Venus.  It represents the scale of the smallest aperture that does not disturb a wave in the actual medium \citep{Hausler2006}. For X-band frequency and $D=10,000$ km, the vertical resolution is about 60 m \citep{Hausler2006}. 

\section{Retrieval of the ionospheric parameters}
\label{Deri}
Radio signal propagating through the Venus' atmosphere and ionosphere is refracted by the surrounding media. Assuming a spherically symmetric atmosphere, each Doppler shift corresponds to a ray path that penetrating the atmosphere down to different depth. If the spatial positions of the spacecraft, the target planet and the ground station are known, the refraction angle and the ray path asymptote altitude can be determined from the Doppler shifts.  The change of the bending angle with respect to the ray path asymptote height can be used to derive the refractivity variation via the Abel integral transform \citep{Fjeldbo1971}. The electron density profile can be computed subsequently, as the refractivity variation is directly related with the local electron density.
 
If the oscillator instabilities are ignored and the effect of Earth's atmosphere is corrected by an atmospheric model, the frequency shift for a one-way radio link due to plasma is given by \citep{Patzold2004}:
\begin{equation}
\Delta{f}=f_r - f_0 =-\frac{f_0}{c}\frac{{\rm d} s}{{\rm d} t}+\frac{1}{c}\frac{{e}^{2}}{8 {\pi}^{2}\epsilon_0 {m}_{e}} \frac{1}{f_{0}}\frac{{\rm d} }{{\rm d} t}\int_{Earth}^{s/c}N_{e} {\rm d} s\ 
\label{eq1}
\end{equation}
where $ {\rm d} s/{\rm d} t $ is the rate of change of the distance between the transmitter and the receiver, $c$ is the speed of light in vacuum, and $f_0$ is the frequency of the signal transmitted from the spacecraft. $f_r$ is the frequency of the signal received at the ground station, $\epsilon_0$ is the permittivity of free space, and $m_e$ is the rest mass of electron. The first term on the right side of equation \ref{eq1} is the classical Doppler shift caused by the relative movement between the spacecraft and the ground station. The second term on the right side is the dispersive effect of ionized media along the ray path from the transmitter to the receiver, which is inversely proportional to $f_0$. The dispersive media includes the interplanetary media, planetary ionosphere and the Earth's ionosphere.

The classical Doppler can be subtracted from the total Doppler shifts by considering the geometrical positions of the spacecraft relative to the ground station. The media effects of Earth's ionosphere can be corrected by an ionospheric model. Then only the unmodeled orbital error and the interplanetary media effect are left in the Doppler residuals, which can further lead to the fluctuations or unrealistic trends in the electron density profiles, especially in the topside where the electron densities are relatively low. The orbital error is proportional to $f_0$, then larger orbital error is expected in the X-band data. 

In addition to the unmodeled orbital errors, the single frequency inversion method can't separate the classical Doppler shifts from the dispersive effects at either S- or X-band alone. This problem can be solved by using the differential Doppler observations, which can be given as $ \Delta{f_{s}}-\frac{3}{11}\Delta{ f_{x}}$, where $\Delta{f_{s}}$ and $\Delta{f_{x}}$ are the observed Doppler shifts at S- and X-band, respectively. The dual-frequency Doppler inversion technique can be referenced to \citet{Zhang2015}.

\begin{figure}[h]
 \centering
  \includegraphics[width=11cm]{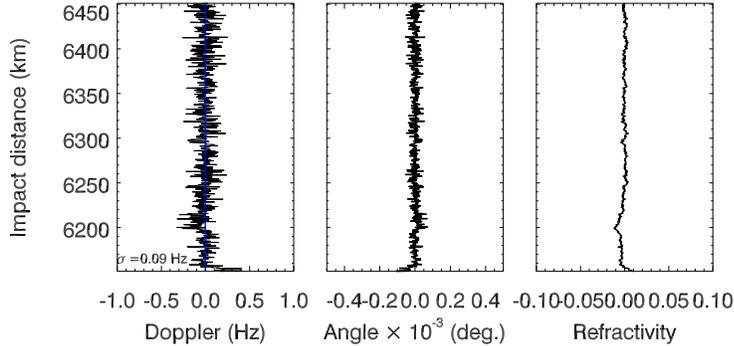}
     \caption{ X-band frequency residuals (left panel), bending angle (middle panel) and refractivity (right panel) variations relative to the impact distance of the RO experiment at Shanghai station. The blue line is the linear regress of the frequency residuals before the baseline correction is applied. A mean radius of 6051.8 km is adopted for Venus.}
  \label{fig1}
\end{figure}

Similar to the Martian RO data processing, a baseline correction is necessary to eliminate the more slowly varying non-Venus ionosphere contributions (e.g., interplanetary space and the Earth's ionosphere) using the data above the reference height \citep{Bird1997}. A suitable reference height will make the retrieved electron density profile around zero both in the 50 - 80 km altitude range and above the ionosphere. The frequency residuals of the X-band RO data observed at Shanghai 25 m station are shown in left panel of Fig. \ref{fig1}, in which a small peak can be found around 147 km. The blue line is the linear regress of the Doppler residuals before the baseline correction is applied. The linear trend is negligible in this observation, but there are conditions where the linear trend is obvious. The standard deviation of the Doppler residuals is 90.0 mHz for 0.1 second integration time and 11.6 mHz for 1 second integration time. This Doppler measurement noise will be used to derive the electron density noise later.
 
 Assuming that the Venus' atmosphere is spherically symmetric, the bending angle $\alpha$ relative to the ray path asymptote altitude $a$ can be solved iteratively from the rays outside the ionosphere to rays at a lower altitude by applying the Bouguer formula \citep{Fjeldbo1971}. The refractivity $N(r)$ can be computed via the Abel transform from the bending angle $\alpha$ through the following equation \citep{Patzold2004}:
\begin{equation}
n(r_1)=\exp (\frac{1}{\pi} \int_{a_1}^{\infty} \frac{\alpha (a)}{\sqrt{a^2-a_1^2}} {\rm d} a)
\label{eq3}
\end{equation}
where $a_1$ represents the impact distance of a ray whose radius of closest approach is $r_1$. The refractivity index $n(r)$ is associated with refractivity $N(r)$ in the form of $N(r)=10^6 \times (n(r)-1)$. The description of the bending angle and impact parameter can be referenced to Fig. 20 of \citet{Fjeldbo1971}.The bending angle and refractivity variations with respect to the impact distance are given in the middle and right panel of Fig. \ref{fig1}, respectively. The integration in Eq. \ref{eq3} can be solved in the following form \citep{Fjeldbo1971}:
\begin{equation}
n(r_1)=\exp (- \frac{1}{\pi} \int_{a_1}^{\infty}  \ln \lbrace \frac{a}{a_1} + \sqrt{(\frac{a}{a_1})^2 -1} \rbrace \frac{{\rm d} \alpha}{{\rm d} a} {\rm d} a)  
\label{eq6}
\end{equation}

 The refractivity as a function of radius is dependent on the local state of the atmosphere and ionosphere \citep{Patzold2004}:
 \begin{equation}
N(r) = K_n N_n(r) - \frac{\kappa_e}{f_0^2} N_e(r)
\label{eq4}
\end{equation}
where $N_n$ is the number density of molecule, $K_n = 1.81\times 10^{-23} {\rm m}^{-3}$ is the mean molecule mass  \citep{VeRaweb,Jenkins1994}, which is dependent on the composition of Venus' atmosphere, $N_e$ is the electron density, $\kappa_e \approx \frac{r_e c^2}{2 \pi} \times 10^6$, $r_e=2.819 \times 10^{-15}$ m is the classical Compton electron radius. In the altitude range of ionosphere, Eq. \ref{eq4} can be simplified as $N(r) \approx - \frac{\kappa_e}{f_0^2} N_e(r)$ . 

The noise of the retrieved electron density profile can be deduced from \citep{Withers2010}:
\begin{equation}
\sigma_{N_e} \approx \frac{4 \pi \sigma_f f c m_e \epsilon_0}{V_{s} e^2} \sqrt[]{\frac{2 \pi H_p}{R_0}}
\label{eq8}
\end{equation}
where the Doppler frequency noise $\sigma_f$ for each observation is computed from the standard deviation of the Doppler residuals in the altitude range of 500 km $-$ 700 km. $V_{s}$ is the relative velocity between spacecraft and ground station , $f$ is the downlink frequency, $H_p =40$ km is the ionospheric scale height (adopted from \citet{Kliore1990}), $R_0=6051.8$ km is the mean radius of Venus.   

\section{Results and discussion}
\subsection{Electron density profile retrieved from the X-band egress RO data at Shanghai station}
\label{Res}
\begin{figure}[h]
 \centering
  \includegraphics[width=7cm]{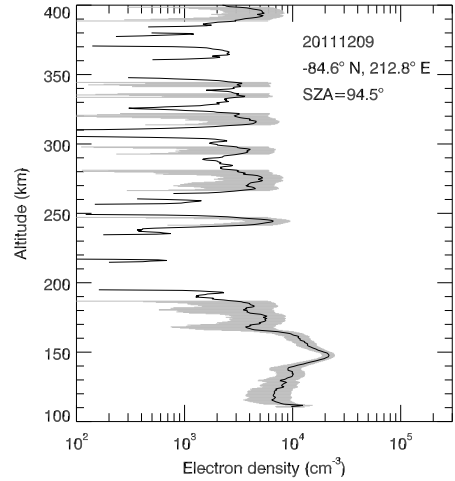}
     \caption{Nightside electron density profile retrieved from the VeRS X-band egress RO data taken in high solar activity, the relevant Doppler residual, bending angle and refractivity data are shown in Fig. \ref{fig1}.The peak altitude is located around 147 km, with a peak density of $2.47 \times 10^4 \; \rm {cm}^{-3}$. The shadowed area is the standard deviation of the electron densities. The related parameters are given in Tab. \ref{tab1}.}
  \label{fig2}
\end{figure}

Due to the lack of S-band observations, single frequency inversion method is adopted to retrieve the electron density profile from the X-band RO tracking data observed by the Shanghai 25 m antenna. The Doppler residual, bending angle and refractivity variations relative to the impact distance are shown in Fig. \ref{fig1}, the retrieved electron density profile is shown in Fig. \ref{fig2}.  As given in Tab. \ref{tab1}, the occultation point of this observation located at (-84.6$^{\circ}$N, 212.8$^{\circ}$E) in the Venus body-fixed coordinate system, and the solar zenith angle is 94.5$^{\circ}$. This is an early morning observation in the high latitude of the southern winter. Solar longitude can represent the seasons on Venus, where 0$^{\circ}$ (180$^{\circ}$) is the vernal (autumnal) equinox, 90$^{\circ}$ (270$^{\circ}$) is the summer (winter) solstice for the northern hemisphere. The reverse applies for the southern hemisphere.

The baseline correction is also applied to the X-band Doppler residuals to move the trend caused by the effects of the interplanetary media and Earth's ionosphere. As explained in Section \ref{Deri}, the unmodeled orbital errors may lead to the unrealistic electron densities, especially in the topside profile where the density is relatively low. Nevertheless, the X-band inversion result can still represent the general state of the local electron densities. As shown in Fig. \ref{fig2}, there is an apparent density peak around 147 km, with a peak density of $2.47 \times 10^4 \rm {cm}^{-3}$. The standard deviation of the density profile is around  $0.26 \times 10^4 \rm {cm}^{-3}$ (10.5\% of the peak density), which is derived from Eq. \ref{eq8} with a $\sigma_f =11.6$ mHz for the 1 Hz Doppler data. Densities below the standard deviation can be treated as noise, which is probably from the measurement noise of the Doppler residuals and the unmodeled orbital errors of the VEX spacecraft.  The densities above 200 km altitude are around the noise level ($0.26 \times 10^4 \rm {cm}^{-3}$), then we can't find an obvious ionopause in this profile.  As this observation is taken during the high solar activity with an average F$_{10.7}$ solar flux of 294 (in unit of $10^{-22} {\rm Wm}^{-2}{\rm Hz}^{-1}$) at Venus, it is reasonable to speculate that the excess of ionospheric thermal pressure relative to the solar wind dynamic pressure raises the ionopause to a much higher altitude. 

\subsection{Electron density profiles retrieved from the VeRa data at New Norcia station}

Part of the VeRa observations at New Norcia station were also processed from August 11, 2006 to June 17, 2007.  The level 2 residual Doppler data are downloaded from the planetary atmospheres data node of NASA PDS (\url{http://pds-atmospheres.nmsu.edu/ve/}). The corresponding Earth, Sun and VEX ephemeris are provided by NAIF SPICE team (\url{http://naif.jpl.nasa.gov/naif/}). This group of datasets contain both ingress and egress occultation data in the single S-, X- or dual-frequency modes.  

\begin{figure}
\begin{center}
   \includegraphics[width=13cm]{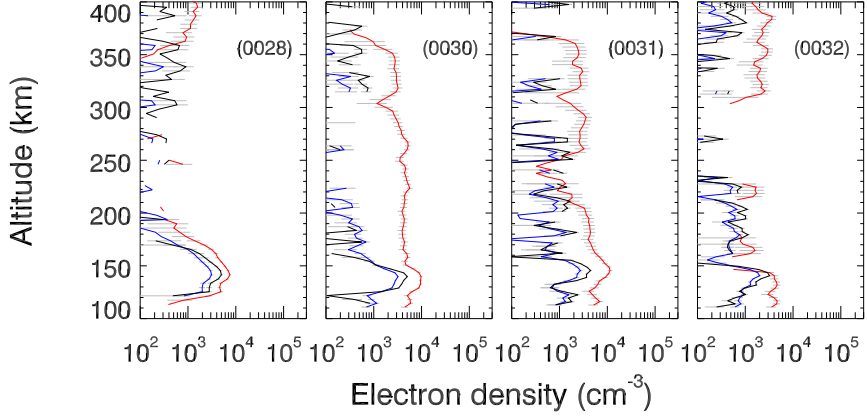}  
\end{center}
   \caption{Electron density profiles retrieved from the S- (blue), X-band (red) and differential (black) ingress RO data for 4 selected datasets of the VeRa nighttime observations taken in intermediate solar activity, with the error bars given as the standard deviation. The related parameters are given in Tab. \ref{tab1}.}
  \label{fig3}
\end{figure}

The single S-, X- and dual-frequency inversion results in ingress mode are shown in Fig. \ref{fig3}, all of them are on the nightside of southern spring. The average F$_{10.7}$ solar radio flux at Venus is around 163 (in unit of $10^{-22} {\rm Wm}^{-2}{\rm Hz}^{-1}$) in this period. From Fig. \ref{fig3}, we can see the electron density profiles retrieved different modes are generally consistent with each other, with the peak density varies from  $3.2 \times 10^{3}$~cm$^{-3}$ to $5.0 \times 10^{4}$~cm$^{-3}$ (see Tab. \ref{tab1}). The percentage of the standard deviation relative to peak electron density ($\sigma_{N_e}/N_m$) varies from 1\% to 8\%, which is much less than that of the observation made at Shanghai station. Meanwhile, $\sigma_{N_e}/N_m$ is larger for nightside profiles compared with dayside data.

As explained in Section \ref{Deri}, profiles retrieved from the differential Doppler observations are more reliable than the single frequency inversion results. Compared with the dual-frequency results (black curves), S-band results (blue curves) can maintain the general shape of the profile but slightly underestimate the peak density.  X-band results  (red curves) overestimate the electron density throughout the profile, especially for the profiles 0030 and 0031. The relatively larger difference between the X-band and the dual-frequency results are mainly due to the unmodeled orbital errors, which is positively related with the radio frequency. 

\begin{figure}[h]
 \centering
  \includegraphics[width=13cm]{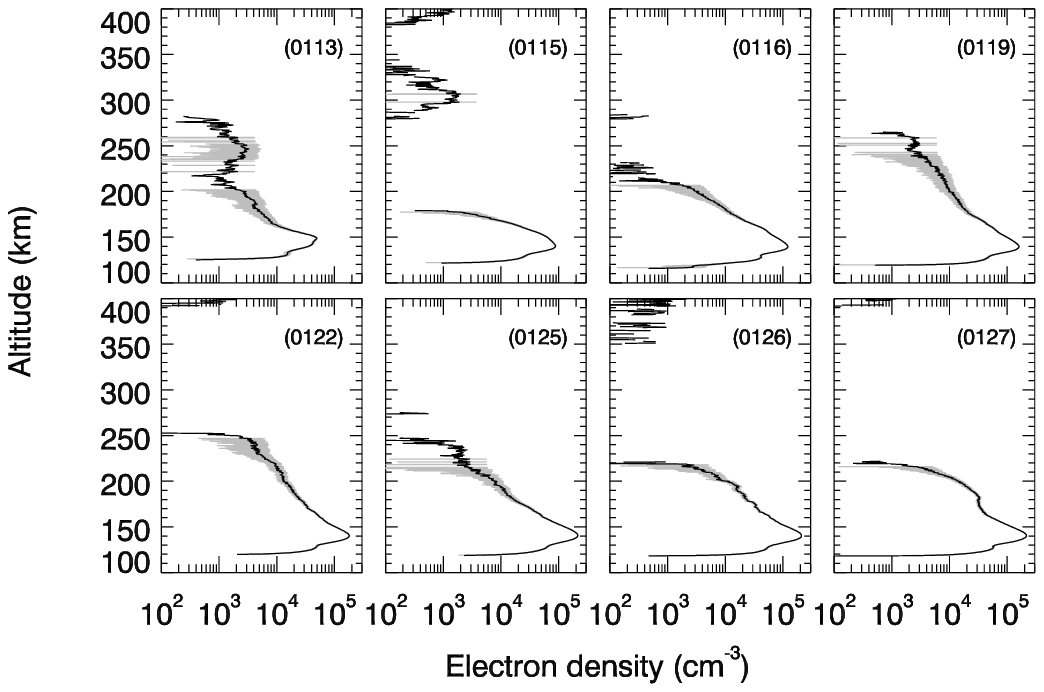}
     \caption{Electron density profiles retrieved from the VeRS X-band ingress observations in intermediate solar activity, with the shadowed area given as the standard deviation. The relevant parameters are given in Tab. \ref{tab1}.}
  \label{fig4}
\end{figure}

Another 8 profiles which are retrieved from the single X-band VeRa observations within 23 days are shown in Fig. \ref{fig4} to show the solar control of the Venus' ionosphere. The local true solar time varies from 1.7 h to 19 h. The peak density of the main layer increases from $0.5 \times 10^{5} {\rm cm}^{-3}$ to $2.1 \times 10^{5} {\rm cm}^{-3}$ with the solar zenith angle (SZA) decreases from 87$^\circ$ to 27$^\circ$  as given in Tab. \ref{tab1}, which clearly shows the solar control of the photochemical layer.  Although the profiles derived from single frequency data may deviate from the real situation in some extent due to the unmodeled orbital errors and interplanetary media effects, we consider here only the relative variations among these profiles.

\begin{figure}[h]
 \centering
  \includegraphics[width=13cm]{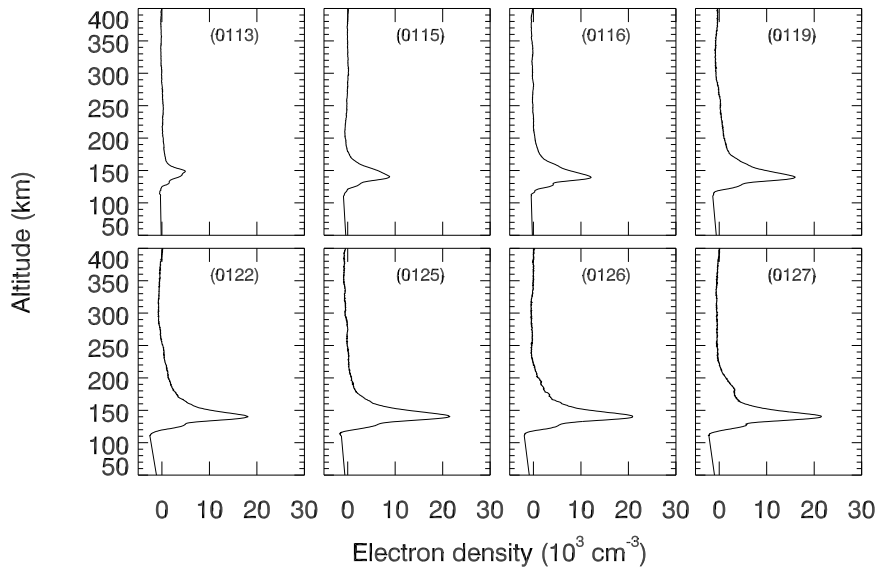}
     \caption{Seem with Fig.\ref{fig4}, but for a linear-scale plot.}
  \label{fig5}
\end{figure}
 \begin{table}[h]
\bc
\begin{minipage}[c]{12cm}
\caption{ Relevant parameters of the profiles shown in Fig. \ref{fig2}, \ref{fig3} and \ref{fig4}.}
\label{tab1}
\end{minipage}
\setlength{\tabcolsep}{1pt}
\small
 \begin{tabular}{ccccccccccccc}
  \hline\noalign{\smallskip}
Profile & Date & DOY& SZA  & Lat & Lon & Ls & Local time & $H_m$ & $N_m$ & $\sigma_{f\_X}$& $\sigma_{N_e}$ &  $\sigma_{N_e}/ N_m$ \\
           & YY/MM/DD & DDD & $^{\circ}$ & $^{\circ}$N & $^{\circ}$E & $^{\circ}$ & h & km & $10^3 $cm$^{-3}$ & mHz & $10^3 $cm$^{-3}$ &   \\
  \hline\noalign{\smallskip}
0001 & 2011/12/09 & 343 & 94.5 & -84.6 & 212.8  &  236.9 & 4.7 &147.1&  24.7&11.6& 2.6 &0.105\\  
0028 & 2006/08/11 & 223 &100.9 & -65.4 & 201.2 &  358.3  &20.0&141.5& 5.0 & 6.9  & 0.4    & 0.081 \\
0030 & 2006/08/16 & 228 & 107.9 & -46.1 & 214.4&  6.3 &  19.9   &139.8& 5.0 &12.3 &0.17 &0.034 \\
0031 & 2006/08/19 & 231 & 111.0 & -33.8 & 222.4 & 11.2 & 19.8  & 145.5& 4.5 & 7.7 &0.24  &0.055 \\
0032 & 2006/08/21 & 233 & 112.5 & -25.1 & 227.7 & 14.4  & 19.7 & 140.6& 3.2 &7.2 &0.11 &0.034\\

0113 & 2007/05/26 & 146 & 88.9 & -87.0 & 356.2 & 101.1 & 19.0 &148.9&  49.2 & 17.5&2.3 & 0.046\\
0115 & 2007/05/28 & 148 & 83.2 & -83.9 & 291.9& 104.4 & 14.3 &140.7& 88.3   & 14.8 &1.8 &0.021\\
0116 & 2007/05/30 & 150 & 77.6 & -78.5 & 282.9 & 107.6 & 13.3 &139.1& 121.3 &18.6&2.2 &0.018\\
0119 & 2007/06/03 & 154 & 66.2 & -67.3 & 285.2 &  114.1 &12.7 &139.8& 159.9& 29.6&3.1 &0.02\\
0122 & 2007/06/07 & 158 & 55.0 & -56.0 & 292.3 & 120.5  & 12.3&140.3&181.0& 28.0& 2.7&0.015\\
0125 & 2007/06/13 & 164 & 38.2 & -38.9 & 305.0& 130.2  & 11.9 &140.1& 214.5& 30.7&2.8 &0.013\\
0126 & 2007/06/15 & 166 & 32.6 & 33.1 & 309.3 &  133.4  & 11.8 &140.4&208.9& 28.2&2.5 &0.012\\
0127 & 2007/06/17 & 168 & 27.1 & 27.2 & 313.8 &  136.7 & 11.7 &140.5&215.2 & 26.1&2.3  &0.011\\
 \noalign{\smallskip}\hline
\end{tabular}
\ec
\tablecomments{0.86\textwidth}{Profile 0001 is the observation made by Shanghai 25 antenna; Ls: solar longitude; Lat and Lon are the latitude and longitude of the occultation footpoint in Venus fixed coordinate system at 100 km altitude. $H_m$ and $N_m$: the peak altitude and peak density of the electron density profile. $\sigma_{f\_X}$: the standard deviation of the Doppler residual at X-band of 1Hz sample frequency; $\sigma_{N_e}$: the standard deviation of the density profile derived from Eq. \ref{eq8}. }
\end{table}

All profiles in Fig. \ref{fig4} have shown the sharp and well-defined peak in a narrow altitude range, and considerable ionizations above the peak altitude except for profile 0115, while the Mars data usually show a broad ionization peak (see Fig. 7(c) of \citet{Kliore1992}). This phenomenon is more obvious in the linear-scale plot given in Fig. \ref{fig5}. The peak density of the nightside profile 0113 is around $0.5 \times 10^{5} {\rm cm}^{-3}$, which is comparable with the peak densities of the profiles in Fig. \ref{fig4}, and much higher than that of the profile in Fig. \ref{fig3}. In contrast, the peak density of the Martian nightside ionosphere is usually below $0.3 \times 10^{5} {\rm cm}^{-3}$. 

The region from 140 km to 180 km is in photochemical equilibrium \citep{Schunk2000}, which is formed by the photoionization of the major neutral components CO$_2$, O$_2$ and O. The altitude above 180 km is the diffusion region and above which a disturbed photodynamical region can be formed by the downward and horizontal plasma flows induced by the solar wind when the dynamic pressure is high \citep{Mahajan1989a}. The diffusion region can extend to much higher altitude if the solar wind pressure is low (see Fig. 1 of \citep{Mahajan1989a}). 

We can find ``bulges" around 180 km altitude in profile 0127, around 250 km in profiles 0113 and 0119 and around 230 km in profiles 0122 and 0125. The bulges  may be caused by the increase of electron temperature similar to the Martian ionosphere\citep{Fox2006}, or just a photodynamcial layer formed due to the pressure exerted by the solar wind as claimed by \citet{Mahajan1989a,Mahajan1989b}.

We can also find density fluctuations on the top of profiles 0113, 0119 and 0125, which may be from the measurement noise \citep{Kliore1992} or the wavelike structures produced by the interaction between the ionosphere and solar wind \citep[and references therein]{Luhmann1991}.  These fluctuations still needs further study, as \citet{Wang2002} stated that ``It does appear though that superposed on these noisy fluctuations, there often are spatial fluctuations present. It is too early to attribute these spatial fluctuations to waves, but they do suggest plasma density variations along the vertical direction.'' 

The average F$_{10.7}$ solar radio flux at Venus is about 153 (in unit of $10^{-22} {\rm Wm}^{-2}{\rm Hz}^{-1}$) during this period.  As the solar activity is in intermediate level, the solar wind pressure may suppress the ionospheric pressure on the dayside, then an ionopause can form in a relatively low altitude in the Venus' ionosphere. If the altitude where the electron density first falls below $5 \times 10^{2}$ cm$^{-3}$ \citep{Kliore1992} is defined as the ionopause, we can clearly find ionopauses in the range of 180 km - 280 km from Fig. \ref{fig4} (around 180 km in profile 0115; around 225 km in profiles 0118, 0126 and 0127; around 250 km in profiles 0119, 0122 and 0125, around 280 km in profile 0113).

As indicated by \citet{Kliore1991} and \citet{Kliore1992}, the ionopause height is generally low for SZAs below 50$^\circ$ regardless of solar activity, and highly variable in the range of 200 km - 1000 km for $55 ^\circ \leq$ SZAs $\leq 90^\circ$ in solar maximum and at times of intermediate conditions, and generally between 200 km and 300 km in solar minimum . As shown in Fig. 10 of \citet{Kliore1991}, the response of the Venus' ionosphere to solar wind dynamic pressure variations is quite constant at solar minimum compared to the profiles in solar maximum and intermediate conditions. \citet{Phillips1988} compared the SZA behavior of the ionopause by different definitions based on the in-situ PVO measurements, from which the ionopause rises from about 350 km in the subsolar region to over 1000 km at a SZA of 120 $^\circ$. 

\section{Conclusion}

Single band inversion and dual-frequency differential Doppler inversion method are used in this paper to retrieve the electron density profiles from the VEX RO data observed by Shanghai 25 m antenna and part of the VeRa observations by New Norcia 35 m antenna. Compared with the X-band data, S-band results agree well with the differential Doppler results in the profile shape, but generally slightly underestimate the peak density.  The discrepancy of the X-band results is mainly due to the unmodeled orbital errors that remain in the Doppler residuals after the geometrical and media Doppler effects are removed. Nevertheless, the X-band data can be used to represent the general state of the Venus' ionosphere, if the S-band data is unavailable. The electron densities of the nightside profiles in Fi.g \ref{fig2} and \ref{fig3} at solar maximum all decrease gradually with the altitude to the noise level, which may indicate that the ionopause is high above 400 km (the upper limit of the figure). The ionopause of profiles in Fig. \ref{fig4} in intermediate solar activity varies from 180 km to 280 km. This result is generally consistent with that given by \citet{Kliore1991} and \citet{Phillips1988}. As the ionosphere of Mars is similar to that of Venus in solar minimum conditions \citep{Kliore1992}, then in most cases, the peak density of the Venus' nightside ionosphere is larger than that of the Martian nightside ionosphere. 

\normalem
\begin{acknowledgements}
This work is supported by the National Natural Science Foundation of China (Grant No. 11103063, 11178008), the national key basic research and development plan (Grant No. 2015CB857101), and partly supported by the Key Laboratory of Planetary Sciences, Chinese Academy of Sciences (Grant No. PSL15\_04). We acknowledge the Venus Express radio science group for publishing the Venus Express radio occultation data online, and the NAIF team for providing the SPICE software. Finally, we are thankful to the anonymous reviewers for greatly improving the quality of the paper.
\end{acknowledgements}
  
\bibliographystyle{raa}
\bibliography{Venus2014}

\begin{thebibliography}{57}
\providecommand{\natexlab}[1]{#1}
\providecommand{\selectlanguage}[1]{\relax}

\bibitem[{Bird et~al.(1997)Bird, Dutta-Roy, Asmar, \& Rebold}]{Bird1997}
Bird, M.~K., Dutta-Roy, R., Asmar, S.~W., \& Rebold, T.~A. 1997, Icarus, 130,
  426

\bibitem[{Brace et~al.(1983)Brace, Elphic, Curtis, \& Russell}]{Brace1983}
Brace, L.~H., Elphic, R.~C., Curtis, S.~A., \& Russell, C.~T. 1983, Geophysical
  Research Letters, 10, 1116

\bibitem[{Brace \& Kliore(1991)}]{Brace1991}
Brace, L.~H., \& Kliore, A.~J. 1991, Space Science Reviews, 55, 81

\bibitem[{Brace et~al.(1980)Brace, Theis, Hoegy et~al.}]{Brace1980}
Brace, L.~H., Theis, R.~F., Hoegy, W.~R., et~al. 1980, Journal of Geophysical
  Research: Space Physics, 85, 7663

\bibitem[{Brace et~al.(1982)Brace, Theis, Mayr, Curtis, \& Luhmann}]{Brace1982}
Brace, L.~H., Theis, R.~F., Mayr, H.~G., Curtis, S.~A., \& Luhmann, J.~G. 1982,
  Journal of Geophysical Research: Space Physics, 87, 199

\bibitem[{Cravens et~al.(1980)Cravens, Gombosi, Kozyra et~al.}]{Cravens1980}
Cravens, T.~E., Gombosi, T.~I., Kozyra, J., et~al. 1980, Journal of Geophysical
  Research: Space Physics (1978–2012), 85, 7778, wiley Online Library

\bibitem[{Cravens et~al.(1981{\natexlab{a}})Cravens, Kliore, Kozyra, \&
  Nagy}]{Cravens1981a}
Cravens, T.~E., Kliore, A.~J., Kozyra, J.~U., \& Nagy, A.~F.
  1981{\natexlab{a}}, Journal of Geophysical Research: Space Physics, 86, 11323

\bibitem[{Cravens et~al.(1981{\natexlab{b}})Cravens, Nagy, \&
  Gombosi}]{Cravens1981b}
Cravens, T.~E., Nagy, A.~F., \& Gombosi, T.~I. 1981{\natexlab{b}}, Advances in
  Space Research, 1, 33

\bibitem[{Fjeldbo et~al.(1975)Fjeldbo, Boris, Donald, \& Taylor}]{Fjeldbo1975}
Fjeldbo, G., Boris, S., Donald, S., \& Taylor, H. 1975, Journal of the
  Atmospheric Sciences, 32, 1232

\bibitem[{Fjeldbo et~al.(1971)Fjeldbo, Kliore, \& R.}]{Fjeldbo1971}
Fjeldbo, G., Kliore, A.~J., \& R., E.~V. 1971, Astronomical Journal, 76

\bibitem[{Fox(1992)}]{Fox1992}
Fox, J.~L. 1992, Planetary and Space Science, 40, 1663

\bibitem[{Fox(2011)}]{Fox2011}
Fox, J.~L. 2011, Icarus, 216, 625

\bibitem[{Fox \& Kasprzak(2007)}]{Fox2007}
Fox, J.~L., \& Kasprzak, W.~T. 2007, Journal of Geophysical Research-Planets,
  112

\bibitem[{Fox \& Yeager(2006)}]{Fox2006}
Fox, J.~L., \& Yeager, K.~E. 2006, Journal of Geophysical Research-Space
  Physics, 111

\bibitem[{Grebowsky \& Curtis(1981)}]{Grebowsky1981}
Grebowsky, J.~M., \& Curtis, S.~A. 1981, Geophysical Research Letters, 8, 1273

\bibitem[{Grebowsky et~al.(1983)Grebowsky, Mayr, Curtis, \&
  Taylor}]{Grebowsky1983}
Grebowsky, J.~M., Mayr, H.~G., Curtis, S.~A., \& Taylor, H.~A. 1983, Journal of
  Geophysical Research: Space Physics, 88, 3005

\bibitem[{H\"{a}usler et~al.(2006{\natexlab{a}})H\"{a}usler, P\"{a}tzold,
  Tyler, Simpson, \& Bird}]{VeRaweb}
H\"{a}usler, B., P\"{a}tzold, M., Tyler, G.~L., Simpson, R.~A., \& Bird, M.~K.
  2006{\natexlab{a}}, Venus Atmospheric, Ionospheric, Surface and
  Interplanetary Radio Wave Propagation Studies with the VeRA Radio-Science
  Experiment, vol. SP-1295 (ESA: ESA Scientific Publication)

\bibitem[{H\"{a}usler et~al.(2006{\natexlab{b}})H\"{a}usler, P\"{a}tzold, Tyler
  et~al.}]{Hausler2006}
H\"{a}usler, B., P\"{a}tzold, M., Tyler, G.~L., et~al. 2006{\natexlab{b}},
  Planetary and Space Science, 54, 1315

\bibitem[{Ivanov-Kholodny et~al.(1979)Ivanov-Kholodny, Kolosov, Savich
  et~al.}]{Kholodny1979}
Ivanov-Kholodny, G.~S., Kolosov, M.~A., Savich, N.~A., et~al. 1979, Icarus, 39,
  209

\bibitem[{Jenkins et~al.(1994)Jenkins, Steffes, \& Hinson}]{Jenkins1994}
Jenkins, J.~M., Steffes, P.~G., \& Hinson, D.~P. 1994, Icarus, 110, 79

\bibitem[{Kim et~al.(1989)Kim, Nagy, Cravens, \& Kliore}]{Kim1989}
Kim, J., Nagy, A.~F., Cravens, T.~E., \& Kliore, A.~J. 1989, Journal of
  Geophysical Research: Space Physics, 94, 11997

\bibitem[{Kliore et~al.(1967)Kliore, Levy, Cain, Fjeldbo, \&
  Rasool}]{Kliore1967}
Kliore, A., Levy, G.~S., Cain, D.~L., Fjeldbo, G., \& Rasool, S.~I. 1967,
  Science, 158, 1683

\bibitem[{Kliore(1992)}]{Kliore1992}
Kliore, A.~J. 1992, Radio Occultation Observations of the Ionospheres of Mars
  and Venus, 265--276 (American Geophysical Union)

\bibitem[{Kliore \& Luhmann(1991)}]{Kliore1991}
Kliore, A.~J., \& Luhmann, J.~G. 1991, Journal of Geophysical Research: Space
  Physics, 96, 21281

\bibitem[{Kliore \& Mullen(1990)}]{Kliore1990}
Kliore, A.~J., \& Mullen, L.~F. 1990, Advances in Space Research, 10, 15

\bibitem[{Knudsen(1992)}]{Knudsen1992}
Knudsen, W.~C. 1992, The Venus Ionosphere from in Situ Measurements, 237--263
  (American Geophysical Union)

\bibitem[{Knudsen et~al.(1979)Knudsen, Spenner, Whitten et~al.}]{Knudsen1979}
Knudsen, W.~C., Spenner, K., Whitten, R.~C., et~al. 1979, Science, 203, 757

\bibitem[{Luhmann \& Bauer(1992)}]{Luhmann1992}
Luhmann, J.~G., \& Bauer, S.~J. 1992, Solar Wind Effects on Atmosphere
  Evolution at Venus and Mars, 417--430 (American Geophysical Union)

\bibitem[{Luhmann \& Cravens(1991)}]{Luhmann1991}
Luhmann, J.~G., \& Cravens, T.~E. 1991, Space Science Reviews, 55, 201

\bibitem[{Luhmann et~al.(1982)Luhmann, Russell, Brace, \& al.}]{Luhmann1982}
Luhmann, J.~G., Russell, C.~T., Brace, L.~H., \& al., e. 1982, Journal of
  Geophysical Research, 87, 9205

\bibitem[{Mahajan \& Mayr(1989)}]{Mahajan1989a}
Mahajan, K.~K., \& Mayr, H.~G. 1989, Geophysical Research Letters, 16, 1477

\bibitem[{Mahajan et~al.(1989)Mahajan, Mayr, Brace, \& Cloutier}]{Mahajan1989b}
Mahajan, K.~K., Mayr, H.~G., Brace, L.~H., \& Cloutier, P.~A. 1989, Geophysical
  Research Letters, 16, 759

\bibitem[{Mahajan \& Oyama(2001)}]{Mahajan2001}
Mahajan, K.~K., \& Oyama, K.~I. 2001, Advances in Space Research, 27, 1863

\bibitem[{Mariner-Stanford-Group(1967)}]{Mariner1967}
Mariner-Stanford-Group 1967, Science, 158, 1678

\bibitem[{Nagy et~al.(1980)Nagy, Cravens, Smith, Taylor, \& Brinton}]{Nagy1980}
Nagy, A.~F., Cravens, T.~E., Smith, S.~G., Taylor, H.~A., \& Brinton, H.~C.
  1980, Journal of Geophysical Research: Space Physics, 85

\bibitem[{P\"{a}tzold et~al.(2007)P\"{a}tzold, H\"{a}usler, Bird
  et~al.}]{Patzold2007}
P\"{a}tzold, M., H\"{a}usler, B., Bird, M.~K., et~al. 2007, Nature, 450, 657

\bibitem[{P\"{a}tzold et~al.(2004)P\"{a}tzold, Neubauer, Carone, \&
  Hagermann}]{Patzold2004}
P\"{a}tzold, M., Neubauer, F.~M., Carone, L., \& Hagermann, A. 2004, MaRS: Mars
  Express Orbiter Radio Science (Eur. Space Agency Spec. Publ. ESA)

\bibitem[{P\"{a}tzold et~al.(2009)P\"{a}tzold, Tellmann, H\"{a}usler
  et~al.}]{Patzold2009}
P\"{a}tzold, M., Tellmann, S., H\"{a}usler, B., et~al. 2009, Geophysical
  research letters, 36

\bibitem[{Peter et~al.(2014)Peter, P\"{a}tzold, Molina-Cuberos
  et~al.}]{Peter2014}
Peter, K., P\"{a}tzold, M., Molina-Cuberos, G., et~al. 2014, Icarus, 233, 66

\bibitem[{Phillips et~al.(1988)Phillips, Luhmann, Knudsen, \&
  Brace}]{Phillips1988}
Phillips, J.~L., Luhmann, J.~G., Knudsen, W.~C., \& Brace, L.~H. 1988, Journal
  of Geophysical Research: Space Physics (1978–2012), 93, 3927, wiley Online
  Library

\bibitem[{Russell et~al.(1980)Russell, Elphic, \& Slavin}]{Russell1980}
Russell, C.~T., Elphic, R.~C., \& Slavin, J.~A. 1980, Journal of Geophysical
  Research: Space Physics, 85, 8319

\bibitem[{Russell et~al.(2006)Russell, Luhmann, \& Strangeway}]{Russell2006}
Russell, C.~T., Luhmann, J.~G., \& Strangeway, R.~J. 2006, Planetary and Space
  Science, 54, 1482

\bibitem[{Russell(1984)}]{Phillips1984}
Russell, J. L. P. J. G. L. C.~T. 1984, Journal of Geophysical Research: Space
  Physics, 89

\bibitem[{Schunk \& Nagy(2000)}]{Schunk2000}
Schunk, R.~W., \& Nagy, A.~F. 2000, Ionospheres - Physics, Plasma Physics, and
  Chemistry (Cambridge, United Kingdom: Cambridge university press)

\bibitem[{Shinagawa(1993)}]{Shinagawa1993}
Shinagawa, H. 1993, Geophysical Research Letters, 20, 2743

\bibitem[{Shinagawa(1996{\natexlab{a}})}]{Shinagawa1996a}
Shinagawa, H. 1996{\natexlab{a}}, Journal of Geophysical Research: Space
  Physics, 101, 26911

\bibitem[{Shinagawa(1996{\natexlab{b}})}]{Shinagawa1996b}
Shinagawa, H. 1996{\natexlab{b}}, Journal of Geophysical Research: Space
  Physics, 101, 26921

\bibitem[{Shinagawa \& Cravens(1988)}]{Shinagawa1988}
Shinagawa, H., \& Cravens, T.~E. 1988, Journal of Geophysical Research: Space
  Physics, 93, 11263

\bibitem[{Svedhem et~al.(2007)Svedhem, Titov, Mccoy et~al.}]{Svedhem2007}
Svedhem, H., Titov, D.~V., Mccoy, D., et~al. 2007, Planetary and Space Science,
  55, 1636

\bibitem[{Taylor et~al.(1980)Taylor, Brinton, Bauer et~al.}]{Taylor1980}
Taylor, H.~A., Brinton, H.~C., Bauer, S.~J., et~al. 1980, Journal of
  Geophysical Research: Space Physics, 85, 7765

\bibitem[{Terada et~al.(2004)Terada, Shinagawa, \& Machida}]{Terada2004}
Terada, N., Shinagawa, H., \& Machida, S. 2004, Advances in Space Research, 33,
  161

\bibitem[{Terada et~al.(2009)Terada, Shinagawa, Tanaka, Murawski, \&
  Terada}]{Terada2009}
Terada, N., Shinagawa, H., Tanaka, T., Murawski, K., \& Terada, K. 2009,
  Journal of Geophysical Research: Space Physics, 114, A09208

\bibitem[{wang \& Nielsen(2002)}]{Wang2002}
wang, J.~S., \& Nielsen, E. 2002, Journal of Geophysical Research, 107

\bibitem[{Withers(2010)}]{Withers2010}
Withers, P. 2010, Advances in Space Research, 46, 58

\bibitem[{Zhang et~al.(1990)Zhang, Luhmann, \& Kliore}]{Zhang1990}
Zhang, M. H.~G., Luhmann, J.~G., \& Kliore, A.~J. 1990, Journal of Geophysical
  Research-Space Physics, 95, 17095

\bibitem[{Zhang et~al.(2015)Zhang, Cui, Guo et~al.}]{Zhang2015}
Zhang, S., Cui, J., Guo, P., et~al. 2015, Advances in Space Research

\bibitem[{Zhang et~al.(2011)Zhang, Ping, \& Han}]{Zhang2011}
Zhang, S.~J., Ping, J.~S., \& Han, T.~T. 2011, Sci China Phys Mech Astron, 54,
  1359

\end{thebibliography}

\end{document}